\documentstyle[11pt,aaspp]{article}
\def\t0{\theta_{\circ}}

\def\be{\begin{equation}}
\def\en{\end{equation}}

\def\msun{M_{\sun}}

\def\kms{\rm \, km \, s^{-1}}

\begin{document}

\title {A disk census for the nearest group of young stars:\\
Mid-infrared observations of the TW Hydrae Association}
\author{Ray Jayawardhana\altaffilmark{1,2,3},
Lee Hartmann\altaffilmark{1},
Giovanni Fazio\altaffilmark{1},\\
R. Scott Fisher\altaffilmark{2,3,4},
Charles M. Telesco\altaffilmark{2,3,4}
and Robert K. Pi\~na\altaffilmark{3,4}}
\altaffiltext{1}{Harvard-Smithsonian Center for Astrophysics, 60 Garden St., Cambridge, MA 02138; Electronic mail: rayjay@cfa.harvard.edu}
\altaffiltext{2} {Visiting Astronomer, Cerro Tololo Interamerican Observatory,
National Optical Astronomy Observatories, which is operated by the Association
of Universities for Research in Astronomy, Inc. (AURA) under cooperative
agreement with the National Science Foundation.}
\altaffiltext{3} {Visiting Astronomer, W.M. Keck Observatory, which is
operated as a scientific partnership among the California Institute of
Technology, the University of California, and the National Aeronautics
and Space Administration. The Observatory was made possible by the
generous financial support of the W.M. Keck Foundation.}
\altaffiltext{4}{Department of Astronomy, University of Florida, Gainesville, FL 32611}

\begin{abstract}
A group of young, active stars in the vicinity of TW Hydrae has recently
been identified as a possible physical association with a common origin.
Given its proximity ($\sim$50 pc), age ($\sim$10 Myr) and abundance of
binary systems, the TW Hya Association is ideally suited to studies of 
diversity and evolution of circumstellar disks. Here we present mid-infrared
observations of 15 candidate members of the group, 11 of which have
no previous flux measurements at wavelengths longer than 2$\mu$m.
We report the discovery of a possible 10$\mu$m excess in 
CD -33$^{\circ}$7795, which 
may be due to a circumstellar disk or a faint, as yet undetected binary 
companion. Of the other stars, only TW Hya, HD 98800, Hen 3-600A, and
HR 4796A -- all of which were detected by IRAS -- show excess thermal 
emission. Our 10$\mu$m flux measurements for the remaining members of 
the Association are consistent with photospheric emission, allowing us
to rule out dusty inner disks. In light of these findings,
we discuss the origin and age of the TW Hya Association as well as
implications for disk evolution timescales. 
\end{abstract}


\section{Introduction}
Circumstellar disks appear to be a natural consequence of the star 
formation process. Observations of young pre-main-sequence (PMS) stars 
show that many of them are surrounded by optically thick disks of 
solar system dimension with masses comparable to or greater than the 
``minimum-mass solar nebula'' of 0.01 $M_{\odot}$ (see Beckwith 1999
for a review). Infrared emission in excess of stellar photospheric fluxes 
provides the most readily measurable signature of such disks. Excesses 
at $\lambda \leq$ 10 $\mu$m are found in $\sim$50\% of the low-mass stars 
in star-forming regions (Strom et al. 1993). 

It has been suggested that circumstellar disks evolve from optically 
thick to optically thin structures in about 10 Myr (Strom et 
al. 1993). That transition may mark the assembly of grains into 
planetesimals, or 
clearing of the disk by planets. Indeed, low-mass debris disks 
have now been imaged around several main sequence stars with ages ranging
from 10 Myr to 1 Gyr (Jayawardhana et al. 1998; Holland et al. 1998; 
Greaves et al. 1998; Koerner et al. 1998; Trilling \& Brown 1998).
However, age estimates for early-type isolated main sequence stars are
highly uncertain. Therefore, the timescale for disk evolution and planet 
formation is still poorly constrained, and may depend critically on the 
presence or absence of a close binary companion.

The recent discovery of a group of young stars associated with TW Hydrae
offers a unique laboratory to study disk evolution and planet formation.
TW Hya itself was first identified as an ``isolated'' T Tauri star by 
Rucinski \& Krautter (1983). Subsequent searches by de la Reza et al. (1989)
and Gregorio-Hetem et al. (1992) found four other young stars in the vicinity
of TW Hya and suggested that the stars may be kinematically associated.
On the basis of strong X-ray emission from all five systems, Kastner et al. 
(1997) concluded that the group forms a physical association at a distance 
of $\sim$50 pc with an age of 20$\pm$10 Myr. (See Jensen, Cohen, 
\& Neuh\"auser 1998 for a different point of view.) Webb et al. (1999) have 
identified five more T Tauri star systems in the same region of the sky as 
candidate members of the ``TW Hya Association'' (TWA), based on the same 
signatures of youth --namely high X-ray flux, large Li abundance, and strong 
chromospheric activity-- and the same proper motion as the original five 
members. Furthermore, they suggest that the wide binary HR 4796, which contains
an A0V star, is also part 
of the Association, even though its {\it Hipparcos} parallactic distance of 
67 pc places it further away than most other members of the group. The three 
other TWA stars with {\it Hipparcos} distances --TW Hya, HD 98800, and TWA 9--
are at 56, 47 and 50 pc, respectively. 

Being the nearest group of young stars, the TW Hya Association is ideally
suited for sensitive disk searches in the mid-infrared. Furthermore, its
estimated age of $\sim$10 Myr provides a strong constraint on disk 
evolution timescales and fills a significant gap in the age sequence between
$\sim$1-Myr-old T Tauri stars in molecular clouds like Taurus-Auriga and
Chamaeleon and the $\sim$50-Myr-old open clusters such as IC 2602 and 
IC 2391. 

Over the past two years, we have conducted mid-infrared observations of
the candidate TWA stars. Our discovery of a spatially-resolved disk around 
HR 4796A and our high-resolution observations of the close binary Hen 3-600 
have already been reported (Jayawardhana et al. 1998, 1999). Here we present 
observations of the other members of the group, including the discovery of 
a possible 10$\mu$m excess in CD -33$^{\circ}$7795, and discuss
implications for the origin and age of the TW Hya Association as well as 
for disk evolution timescales.  

\section{Observations}
During three observing runs in 1998 and 1999, we have obtained mid-infrared 
images of candidate 
members of the TW Hya Association using the OSCIR instrument on the 
4-meter Blanco telescope at Cerro Tololo Interamerican Observatory (CTIO) 
and the 10-meter Keck II telescope. The log of our observations is 
given in Table 1. OSCIR is a mid-infrared imager/spectrometer built at 
the University of Florida, using a 128$\times$128 Si:As Blocked Impurity 
Band (BIB) detector developed by Boeing. Additional information on  OSCIR 
is available on the Internet at www.astro.ufl.edu/iag/.

On the CTIO 4-m telescope, OSCIR has a plate scale of 0.183''/pixel, which 
gives a field of view of 23''$\times$23''. Our observations were made using 
the standard chop/nod technique with a chop frequency of 5 Hz and a throw of 
23'' in declination. On Keck II, its plate scale is 0.062''/pixel, providing 
a 7.9''$\times$7.9'' field of view. Here we used a chop frequency of 4 Hz and 
a throw of 8''. Images were obtained in the N(10.8 $\mu$m) band for the 
entire sample, and in the IHW18(18.2 $\mu$m) band for a few bright targets.

\section{Results}
In Table 2, we present the measured 10$\mu$m fluxes for the entire sample,
and compare them to the expected photospheric fluxes, assuming {\it K - N}=0
for all late-type stars; For HR 4796A, we used {\it K - N}=-0.03, as given 
by Kenyon \& Hartmann (1995) for an A0V star.

Among the candidate TWA stars, only TW Hya, HD 98800, Hen 3-600A and 
HR 4796A -- all of which were first detected by {\it IRAS} -- show
significant excess at mid-infrared wavelengths. We have detected a modest
10$\mu$m excess in CD -33$^{\circ}$7795 for the first time (see below).
None of the other late-type stars have excess, suggesting that they do not
harbor dusty inner disks. 

\noindent
{\it Comments on individual objects}

{\bf TW Hydrae}
The spectral energy distribution (SED) of TW Hya from near- to far-infrared
wavelengths, including our flux measurements at 10.8$\mu$m and 18.2$\mu$m,
is shown in Figure 1a. The excess at $\lambda \geq$ 20$\mu$m is unusually
strong compared to the median of classical T Tauri stars in Taurus (solid 
line in Figure 1a). It is worth noting that TW Hya also has a large H$\alpha$
equivalent width of -220\AA, consistent with an actively accreting disk.

{\bf CD -33$^{\circ}$7795}
We measure a 10$\mu$m flux of 96$\pm$9 mJy for CD -33$^{\circ}$7795,
somewhat above its estimated photospheric emission of 70$\pm$5 mJy. This 
modest excess, at a level of 2.6$\sigma$, is well below what is expected for 
an optically thick inner disk (Figure 1b). One possibility is that our
assumption {\it K - N}=0 is not correct; if {\it K - N}$\sim$0.3, there
would be no 10$\mu$m excess. However, we note that using {\it K - N}=0
gives good agreement with measured fluxes for other stars of similar spectral
type in the sample. 

If the excess is real, it could be due to an optically thin disk or a 
faint, as yet undetected stellar companion. We note that CD -33$^{\circ}$7795
may be a spectroscopic binary according to Webb et al. (1999). Furthermore, 
Lowrance et al. (1999) have reported the discovery of a possible brown dwarf 
companion 2'', or $\sim$100 AU, from CD -33$^{\circ}$7795. Our 10$\mu$m flux 
measurement is within an aperture of 1'' radius (seeing$\approx$0.7''), and 
thus should not include a contribution from this brown dwarf. In any case, 
to account for the observed 10$\mu$m excess, the brown dwarf ($K=11.5$) would 
have to have {\it K - N}=3.6!

{\bf HR 4796B}
We placed an upper limit of 23 mJy to the 10$\mu$m emission from HR 4796B
from CTIO data and have now detected its photosphere at 16$\pm$2 mJy at Keck. 
This result is of particular interest because the age of HR 4796B is
fairly well established. Using the {\it Hipparcos} distance of 67 pc
to HR 4796A and D'Antona and Mazzitelli (1994) evolutionary tracks, it is 
possible to estimate an age of $8 \pm 3$~Myr for B which is an M2.5 star
(Jayawardhana et al. 1998).  This age is consistent with the upper bound
provided by the measurement of the strong Li absorption line at 6708 \AA~
(Stauffer et al. 1995). The lack of 10$\mu$m excess in this object suggests
that HR 4796B does not have a dusty inner disk. Unfortunately, we cannot 
determine at present whether it originally had an optically thick disk, 
which has since depleted, or whether it formed without a disk (like some 
50\% of T Tauri stars appear to be). If future far-infrared and sub-millimeter
observations find evidence for an outer disk around HR 4796B, that would 
argue for rapid evolution of the inner disk, either through coagulation of 
dust or accretion on to the central star.

\section{Discussion}
\subsection{Origin and age of the TW Hya Association}
Whether the TWA stars are physically related in origin is a matter
of controversy.  Kastner et al. (1997) and Webb et al. (1999) argue that
this is an unusual grouping of relatively young (10 Myr old) stars
unlikely to be a chance coincidence, while Jensen et al. (1998) argue 
that the proper motions of three primary members -- TW Hya, HD 98800,
and CD -36$^{\circ}$7429 -- are inconsistent with them having formed
together 10 Myr ago. Confusing the issue further, TW Hya has all the 
characteristics of an actively accreting T Tauri star (Rucinski \& Krautter 
1983; de la Reza et al. 1989; Gregorio-Hetem et al. 1992), typical of much 
younger systems (ages $<$ 3 Myr), while most of the other members show weak 
or no H$\alpha$ emission (Kastner et al. 1997; Webb et al. 1999), and, as 
we have shown here, little or no infrared excess emission from dusty disks.  
 
Our data do not bear directly on the question of the physical connection
between the stars in this association, but a few comments can be made about 
membership determinations.  The first question to be addressed is
whether the TWA members really constitute a special and unlikely 
concentration of objects.  Webb et al. (1999) argue for this conclusion, 
based on three
points: first, essentially that there are few or no classical T Tauri stars 
known outside of clouds; second, that X-ray surveys do not show similar 
numbers of 10 Myr old stars; and third, that there is no evidence for 10-100 
Myr old populations in the solar neighborhood.  The first argument is not 
very strong because there is only one bona fide Classical T Tauri star in the 
TW Hya group and so statistics are poor. The third argument appears to be 
inconsistent with the results of Brice\~no et al. (1997), who showed that
many of the X-ray bright low-mass stars in the ROSAT All Sky Survey (RASS) 
have ages of 50-100 Myr.  The second argument also has problems; as
Mart\'in \& Magazz\`u (1999) showed from a study of Li equivalent widths in 
RASS-selected stars in the direction of Taurus, while most of the systems 
are likely to be 50-100 Myr old, a modest fraction ($\sim 20$\%) of these 
objects may indeed be $\sim 10$~Myr old.

Another way to address the question of the overdensity is to use estimates of 
the expected average birthrate in the solar neighborhood.  Using the results 
of Miller \& Scalo (1979) for a constant birthrate and the age parameter 
$T_o = 12 \times 10^9$~yr, and their form for the initial mass function, 
we predict that the number of stars formed per year between $0.8 \msun$ 
and $0.3 \msun$ in the solar neighborhood is  $\sim 2.1 \times
10^{-9}$~stars~yr$^{-1}$~pc$^{-2}$.  Thus, within a radius
of 75~pc from the Sun, there should be approximately 370 stars with ages 
$\le 10$~Myr. If these stars were distributed uniformly across the sky 
within a band of $\pm 45^{\circ}$ from the galactic equator (i.e., an 
effective scale height of approximately 70 pc), the surface density of such 
objects would then be $\sim 1.3 \times 10^{-2}$ stars per square degree.
The main body of the TW Hya Association identified by Webb et al. (1999) 
spans a range of 13 degress in declination by about 19 degrees in right 
ascension. Thus in these 245 square degrees one should expect on average 3.1 
stars in the 1-10 Myr age range, whereas Webb et al. identify 11 such objects. 
If one includes HR 4796 and TWA 10, as suggested by Webb et al., the
total number of observed objects increases to 14 but the number of
predicted randomly-distributed objects in the larger 17 degree by 28
degree region also goes up to 6.
 
Thus, while the above calculations suggest that the TW Hya Association 
is {\it probably} a significant enhancement in the local density of young 
stars, the possibility that this is a chance alignment cannot be completely 
ruled out.  The predicted average density depends upon parameter choices that 
are not certain, such as the precise volume that is being sampled and the 
actual ages of the stars (changes of a factor of two in these properties 
strongly affect the apparent significance of the grouping).  While it is 
probably not appropriate to use an averaged birthrate for such young stars, 
note that for typical space velocities of $\sim 5 \kms$ (Hoff et al. 1998), 
groups of age 10-20 Myr can overlap if they originated 50-100 pc apart.
 
Regardless of the physical association of these objects (which really only 
is used to support the application of the {\it Hipparcos} distances of the 
main objects to all suggested members), the lack of Li depletion indicates 
that these stars cannot be much older than $\sim$10 Myr. Whatever molecular 
cloud(s) these stars formed in, the absence of relatively nearby 
molecular gas is most easily explained if the natal clouds disperse
in $\le 10$~Myr (Hoff et al. 1998), consistent with the general absence 
of 3-10 Myr old stars in molecular clouds such as Taurus-Auriga (e.g., 
Brice\~no et al. 1997), rather than by requiring very high space velocities 
to move stars from present-day clouds (Soderblom et al. 1998).

\subsection{Implications for disk evolution timescales}
Our mid-infrared observations show that many of the stars in the TW Hya 
Association have little or no disk emission at 10$\mu$m. Even among the 
five stellar systems with 10$\mu$m excesses, most show some evidence 
of inner disk evolution. The disk around the A0 star HR 4796A has an
$r \approx 60$ AU central hole in mid-infrared images (Jayawardhana et al. 
1998; Koerner et al. 1998). The SEDs of HD 98800 and Hen 3-600A also suggest
possible inner disk holes (Jayawardhana et al. 1999). The excess we report 
here for CD -33$^{\circ}$7795 is modest, and could well be due to a faint
companion. Only TW Hya appears to harbor an optically thick, actively
accreting disk of the kind observed in $\sim$1-Myr-old classical
T Tauri stars; it is the only one with a large H$\alpha$ equivalent
width (-220 \AA). It would be of great interest to further constrain TW
Hya's SED with flux measurements at wavelengths between 2-10 $\mu$m. 

If most TWA stars are $\leq$10 Myr old, the above results suggest that 
their inner disks have already depleted either through coagulation of dust 
or accretion on to the central star.  The fact that only one (TW Hya) out of 
16 entries in Table 2 shows classical T Tauri characteristics --compared to
$\sim$50\% of $\sim$1-Myr-old stars in star-forming regions-- argues for
rapid evolution of inner disks in pre-main-sequence stars. Observations at 
far-infrared and sub-millimeter wavelengths may reveal whether most TWA stars 
still retain their outer disks. These stars are also ideal targets for 
sensitive brown dwarf and planet searches with the Space Interferometry 
Mission (SIM) and the proposed Terrestrial Planet Finder (TPF).

\bigskip
We wish to thank the staff of CTIO and Keck Observatory for their 
outstanding support. The research at CfA was supported by NASA 
grant NAG5-4282 and the Smithsonian Institution. The research at the 
University of Florida was supported by NASA, NSF, and the University of 
Florida.

\newpage
\begin{table}
\begin{scriptsize}
\begin{center}
\renewcommand{\arraystretch}{1.2}
\begin{tabular}{lccccc}
\multicolumn{6}{c}{\scriptsize TABLE 1}\\
\multicolumn{6}{c}{\scriptsize LOG OF OBSERVATIONS}\\
\hline
\hline
UT Date & Telescope & Target & Filter & On-source integration & Flux standards\\
\hline
1998 Mar 18 & CTIO 4m & CD -29$^{\circ}$8887 & N & 600 sec & $\lambda$
Vel, $\gamma$ Cru\\
1998 Mar 18 & CTIO 4m & CD -33$^{\circ}$7795 & N & 600 sec & $\lambda$
Vel, $\gamma$ Cru\\
1998 Mar 18 & CTIO 4m & HR 4796A\&B & N & 1800 sec& $\lambda$ Vel, $\gamma$ Cru\\
1998 Mar 18 & CTIO 4m & HR 4796A\&B & IHW18 & 1800 sec& $\lambda$ Vel, $\gamma$ Cru\\
1999 Feb 23 & CTIO 4m & TW Hya  & N & 600 sec & $\lambda$ Vel, $\gamma$ Cru\\
1999 Feb 23 & CTIO 4m & TW Hya& IHW18& 1200 sec & $\lambda$ Vel, $\gamma$ Cru\\
1999 Feb 23 & CTIO 4m & Hen 3-600A\&B& N & 600 sec & $\lambda$ Vel, $\gamma$ Cru\\
1999 Feb 23 & CTIO 4m & Hen 3-600A\&B& IHW18 & 600 sec & $\lambda$ Vel, $\gamma$ Cru\\
1999 Feb 26 & CTIO 4m & TWA 7 & N & 400 sec & $\gamma$ Cru, $\alpha$ CMa\\
1999 May 2  & Keck II & TWA 8A & N & 85 sec & $\mu$ UMa, $\alpha$ Boo\\
1999 May 2  & Keck II & TWA 8B & N & 300 sec & $\mu$ UMa, $\alpha$ Boo\\
1999 May 2  & Keck II & TWA 10 & N & 300 sec & $\mu$ UMa, $\alpha$ Boo\\
1999 May 3  & Keck II & CD -29$^{\circ}$8887& N & 150 sec & $\mu$ UMa,
$\alpha$ Boo\\
1999 May 3  & Keck II & TWA 6 & N & 300 sec & $\mu$ UMa, $\alpha$ Boo\\
1999 May 3  & Keck II & TWA 9A\&B & N & 600 sec & $\mu$ UMa, $\alpha$ Boo\\
1999 May 3  & Keck II & HR 4796B & N & 150 sec & $\mu$ UMa, $\alpha$ Boo\\
\hline
\end{tabular}
\end{center}
\end{scriptsize}
\end{table}

\newpage
\begin{table}
\begin{scriptsize}
\begin{center}
\renewcommand{\arraystretch}{1.2}
\begin{tabular}{lccccc}
\multicolumn{6}{c}{\scriptsize TABLE 2}\\
\multicolumn{6}{c}{\scriptsize N-BAND FLUX MEASUREMENTS}\\
\hline
\hline
TWA Number & Common name & Spectral type & Photospheric N & Measured N & Source\\
 & & & (mJy) & (mJy) & \\
\hline
1  & TW Hya     & K7    & 43 & 580$\pm$58& CTIO\\
2A & CD -29$^{\circ}$8887& M0.5  & 50 & 49$\pm$2 & 
Keck/CTIO\tablenotemark{a}\\
2B & ...        & M2    & 24 & 31$\pm$4 & Keck/CTIO\tablenotemark{a}\\
3A & Hen 3-600A & M3    & 46 & 900$\pm$90& CTIO\\
3B & Hen 3-600B & M3.5  & 29 & $<$50& CTIO\\
4  & HD 98800   & K5    & 208& 1400$\pm$70& Gehrz et al.(1999)\\
5A & CD -33$^{\circ}$7795& M1.5  & 70 & 96$\pm$9 & CTIO\\
6  & ...        & K7    & 24 & 22$\pm$2 & Keck\\
7  & ...        & M1    & 66 & 62$\pm$6& CTIO\\
8A & ...        & M2    & 40 & 41$\pm$2 & Keck\\
8B & ...        & M5    & 9  & 10$\pm$2& Keck\\
9A & CD -36$^{\circ}$7429& K5    & 31 & 26$\pm$2 & Keck\\
9B & ...        & M1    & 8  & 5$\pm$2  & Keck\\
10 & ...        & M2.5  & 20 & 24$\pm$2 & Keck\\
11A& HR 4796A   & A0    & 176& 244$\pm$25& CTIO\\
11B& HR 4796B   & M2.5  & 17 & 16$\pm$2 & Keck/CTIO\tablenotemark{b}\\
\hline
\end{tabular}
\end{center}
\tablenotetext{a}{\scriptsize This 0.6'' binary was not resolved at CTIO. 
However, the total flux we  measured at CTIO (82$\pm$8 mJy) agrees 
extremely well with the sum of fluxes measured at Keck (80 mJy) }
\tablenotetext{b}{\scriptsize Our Keck flux measurement of 16$\pm$2 mJy is 
consistent with the CTIO upper limit of 23 mJy.}
\end{scriptsize}
\end{table}

\newpage

\centerline{\bf Figure Caption}

\bigskip
\bigskip

Figure 1. Composite spectral energy distribution (SED) of (a) TW Hydrae
and (b) CD -33$^{\circ}$7795. The solid line in each plot is the median
SED for Taurus CTTS, normalized at H (from D'Alessio et al. 1999), and
the dashed lines show the quartile fluxes to provide some idea of the
range of observed CTTS fluxes.

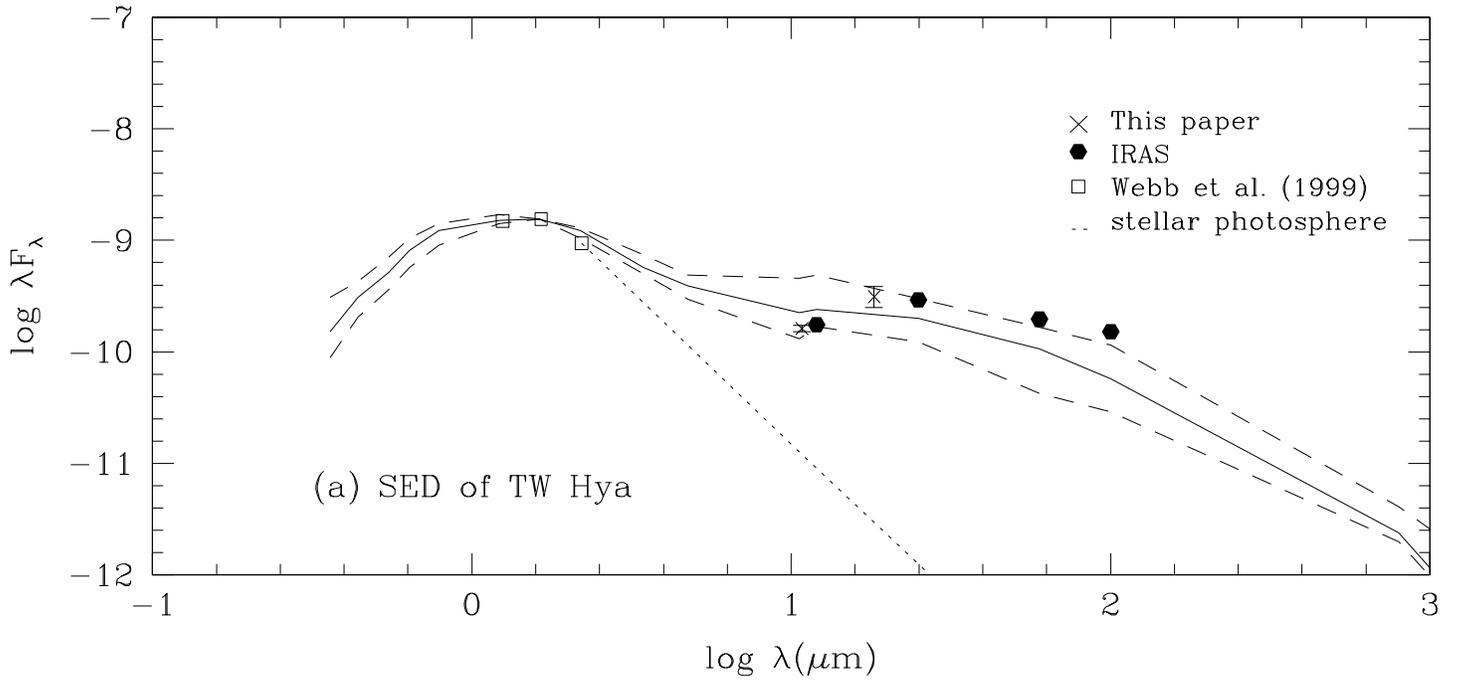
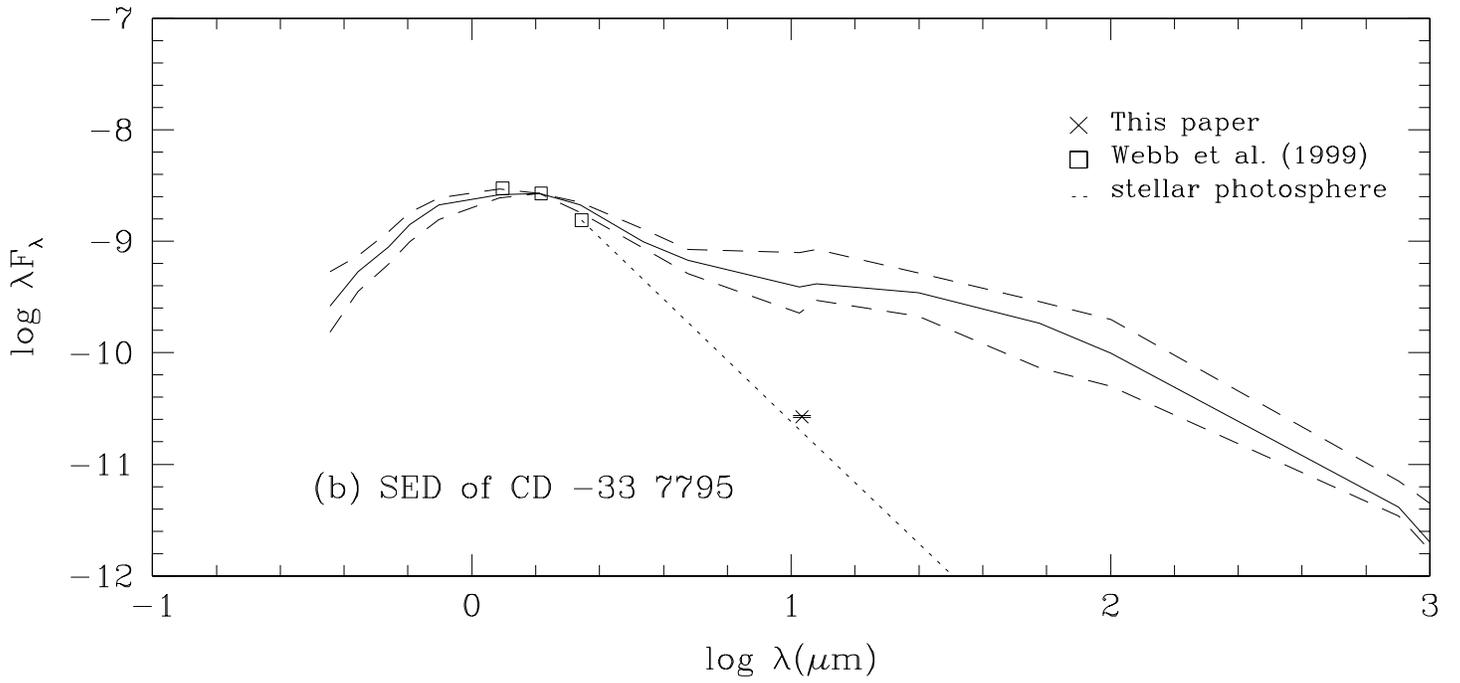
\end{document}